\documentclass[twocolumn,aps,showpacs,floatfix,superscriptaddress]{revtex4}
\usepackage{amsmath,amssymb,eucal,graphicx}
\begin{document}
\title{Scaling in Tournaments}
\author{E.~Ben-Naim}
\affiliation{Theoretical Division and Center for Nonlinear Studies,
Los Alamos National Laboratory, Los Alamos, New Mexico 87545 USA}
\author{S.~Redner}
\affiliation{Department of Physics, Boston University, Boston,
Massachusetts 02215 USA}
\author{F.~Vazquez}
\affiliation{Theoretical Division and Center for Nonlinear Studies,
Los Alamos National Laboratory, Los Alamos, New Mexico 87545 USA}
\affiliation{Department of Physics, Boston University, Boston,
Massachusetts 02215 USA}
\begin{abstract}
We study a stochastic process that mimics single-game elimination
  tournaments. In our model, the outcome of each match is stochastic:
  the weaker player wins with upset probability $q\leq 1/2$, and the
  stronger player wins with probability $1-q$. The loser is
  eliminated.  Extremal statistics of the initial
distribution of player strengths governs the tournament outcome. For
a uniform initial distribution of strengths, the rank of the winner,
$x_*$,  decays algebraically with the number of players, $N$, as
$x_*\sim N^{-\beta}$.  Different decay exponents are found
analytically for  sequential dynamics, \hbox{$\beta_{\rm
seq}=1-2q$}, and parallel dynamics, \hbox{$\beta_{\rm
par}=1+\frac{\ln (1-q)}{\ln 2}$}.  The  distribution of player
strengths becomes self-similar in the long time limit with an
algebraic tail. Our theory successfully describes statistics of the
US college basketball national championship tournament.
\end{abstract}
\maketitle

A wide variety of processes in nature and society involve
competition. In animal societies, competition is responsible for
social differentiation and the emergence of social strata.
Competition is also ubiquitous in human society: auctions, election
of public officials, city plans, grant awards, and sports involve
competition.  Minimalist, physics-based competition processes have
been recently developed to model relevant competitive phenomena such
as wealth distributions \cite{ikr,dy,fs}, auctions \cite{dr,bk,bkk},
social dynamics \cite{btd,br,bvr,msk}, games \cite{truels}, and sports
leagues \cite{bvr1}.  In physics, competition also underlies phase
ordering kinetics, in which large domains grow at the expense of
small domains that eventually are eliminated \cite{gss,ajb}.

In this study, we investigate $N$-player tournaments with head-to-head
matches.  The winner of each match remains in the tournament while the
loser is eliminated.  At the end of a tournament, a single undefeated
player, the tournament winner, remains.  Each player is endowed with a
fixed intrinsic strength $x\geq 0$ that is drawn from a normalized
distribution $f_0(x)$.  We define strength so that smaller $x$
corresponds to a stronger player and we henceforth refer to this
strength measure as ``rank''.

The result of competition is stochastic: in each match the weaker
player wins with the upset probability \hbox{$q\leq 1/2$} and the
stronger player wins with probability \hbox{$p=1-q$}. Schematically,
when two players with ranks $x_1$ and $x_2$ compete, assuming
$x_1<x_2$, the outcome is:
\begin{equation}
\label{rule}
(x_1,x_2)\to
\begin{cases}
x_1&{\rm with\ probability\ }1-q;\\
x_2&{\rm with\ probability\ }q.
\end{cases}
\end{equation}
For $q=0$, the best player is always victorious, while for $q=1/2$,
game outcomes are completely random. We are interested in the
evolution of the rank distribution, as well as the rank of the
tournament winner.

We find that the rank of the winner, $x_*$, decays algebraically with
the number of players $N$ as
\begin{equation}
\label{xn-scaling}
x_*\sim N^{-\beta}
\end{equation}
with the exponent $\beta\equiv \beta(q)$ a function of the upset
probability.  When the ranks of the tournament players are uniformly
distributed, we find different values for sequential and parallel
dynamics: $\beta_{\rm seq}=1-2q$ and $\beta_{\rm par}=1+\frac{\ln
(1-q)}{\ln 2}$.  Moreover, the rank distribution becomes
asymptotically self-similar and has a power-law tail. We also extend
these results to arbitrary initial distributions.  The extreme of
this distribution governs statistical properties of the rank of the
ultimate winner.

\noindent{\bf Sequential Dynamics.} We formulate the competition
process by assuming that each pair of players compete at a constant
rate. In this formulation, games are held sequentially, and players
are eliminated from the tournament one at a time. The fraction of
players remaining in the competition at time $t$, $c(t)$, decays
according to
\begin{equation}
\label{c-eq}
\frac{dc}{dt}=-c^2.
\end{equation}
Solving this equation subject to the initial condition $c(0)=1$, the
surviving fraction is
\begin{equation}
\label{n-sol}
c(t)=(1+t)^{-1}.
\end{equation}
The tournament ends with a single player and this occurs at time
$t_*$, that can be estimated from $c(t_*)\sim N^{-1}$. Therefore the
time to complete the competition scales linearly with the number of
players $t_*\sim N$.

Let $f(x,t)\,dx$ be the fraction of {\it remaining} players with rank
in the range $(x,x+dx)$ at time $t$.  The density $f(x,t)$ obeys the
nonlinear integro-differential equation
\begin{equation}
\label{f-eq}
\frac{\partial f(x)}{\partial t}=
-2p\,f(x)\int_0^x dy f(y)\,
-2q\,f(x)\int_x^{\infty} dy f(y).
\end{equation}
The first term accounts for games where the favorite wins and the
second term for games where the underdog wins. The initial condition
is $f(x,0)=f_0(x)$ with \hbox{$\int dx f_0(x)=1$}. Integrating
(\ref{f-eq}), the total fraction of remaining players,
\hbox{$c(t)=\int dx f(x,t)$}, indeed decays according to
(\ref{c-eq}). We note that this master equation is exact in the limit
of an infinite number of players and applicable only as long as the
fraction of remaining players is finite.

The rank distribution can be obtained by introducing the cumulative
distribution $F(x)$, defined as the fraction of players with rank smaller
than $x$,
\begin{equation}
\label{F-def}
F(x)=\int_0^x dy f(y).
\end{equation}
The distribution of player ranks is obtained from the cumulative
distribution by differentiation, \hbox{$f(x)=dF(x)/dx$}.  By
integrating the master equation (\ref{f-eq}), the cumulative
distribution obeys the closed nonlinear  equation
\begin{equation}
\label{F-eq}
\frac{\partial F}{\partial t}=(2q-1)F^2-2qcF.
\end{equation}
The initial condition is \hbox{$F(x,0)=F_0(x)=\int_0^x dy f_0(y)$}.
Substituting \hbox{$H(x)=1/F(x)$}, we transform (\ref{F-eq}) to the
linear equation
\begin{equation}
\frac{\partial H}{\partial t}=(1-2q)+2qcH.
\end{equation}
Integrating this equation with respect to time, we find
\hbox{$H(x)=[H_0(x)-1](1+t)^{2q}+(1+t)$}. Substituting the initial
condition $H_0(x)=1/F_0(x)$, we obtain the cumulative rank
distribution
\begin{equation}
\label{F-sol}
F(x,t)=\frac{F_0(x)}{[1-F_0(x)](1+t)^{2q}+F_0(x)(1+t)}.
\end{equation}
From this, the actual density of player rank is obtained by
differentiation
\begin{equation}
\label{f-sol}
f(x,t)=\frac{f_0(x)(1+t)^{2q}}
{\left[(1-F_0(x))(1+t)^{2q}+F_0(x)(1+t)\right]^2}.
\end{equation}
Notice that when the game outcome is random, $q=1/2$, the normalized
distribution of rank does not evolve with time as $f(x,t)/c(t) =
f_0(x)$.

\noindent{\bf Uniform Initial Distribution.}  Consider first the
special case of a uniform initial distribution, $f_0(x)=1$ for
$0\leq x \leq 1$, and deterministic games, $q=0$. Then the initial
cumulative distribution is $F_0(x)=x$ for $x\leq 1$ and $F_0(x)=1$
for $x\geq 1$.  The time-dependent cumulative distribution
(\ref{F-sol}) is
\begin{equation}
\label{F-sol-uniform}
F(x,t)=\frac{x}{1+xt}
\end{equation}
for $x\leq 1$ and $F(x,t)=c(t)$ for $x\geq 1$. Similarly, the rank
distribution itself is $f(x,t)=(1+xt)^{-2}$ for \hbox{$0\leq x\leq
1$}. As expected, weaker players are more likely to be eliminated as
the tournament proceeds and the remaining field becomes stronger.
Quantitatively, the average rank of surviving players,
\hbox{$\langle x\rangle=\int dx\, x f(x)/\int dx f(x)$}, is
\begin{equation}
\label{xav}
\langle x\rangle = t^{-2}\left[\,(1+t)\ln (1+t)-t\,\right].
\end{equation}
Therefore, the average rank asymptotically decays with time, $\langle
x\rangle \simeq t^{-1}\ln t$.

We can write the cumulative distribution in the scaling form
$F(x,t)\to t^{-1}\Phi(xt)$, by multiplying and dividing
(\ref{F-sol-uniform}) by time.  Here, the scaling function is
\hbox{$\Phi(z)=\frac{z}{1+z}$}, which approaches unity $\Phi(z)\to 1$
when $z\to\infty$, consistent with total density decay $c\simeq
t^{-1}$.  In the long time limit, the cumulative distribution retains
the same shape as the initial distribution, $\Phi(z)\simeq z$, for
$z\ll 1$. The scaling variable $z=xt$ indicates that players with rank
larger than the characteristic rank $x\sim t^{-1}$ are eliminated from
the tournament.

Let us generalize these results to arbitrary $q$. In this case, the
cumulative distribution is
\begin{equation}
F(x,t)=\frac{x}{(1-x)(1+t)^{2q}+x(1+t)},
\end{equation}
for $x\leq 1$ and $F(x,t)=c(t)$ otherwise. In the long time limit, we may
replace $1+t$ with $t$, and also replace $1-x$ with $1$, since the rank
decays with time. Then the cumulative distribution approaches the scaling
form
\begin{equation}
\label{Fxt-scaling}
F(x,t)\to t^{-1}\Phi\left(x\,t^{1-2q}\right).
\end{equation}
The scaling function remains as above
\begin{equation}
\label{phiz}
\Phi(z)=\frac{z}{1+z}.
\end{equation}
The scaling form (\ref{Fxt-scaling}) implies that the typical rank
decays algebraically with time
\begin{equation}
\label{xt-scaling}
x\sim t^{-(1-2q)}.
\end{equation}
Interestingly, the exponent governing this decay depends on the upset
probability.  The larger the upset probability, the smaller the decay
exponent.  Thus weaker players can persist in a tournament when $q$
approaches 1/2.  For completely random games, $q=1/2$, the exponent
vanishes and the strength of the typical surviving player does not
change with time.

A similar scaling law characterizes the rank of the tournament
winner. From (\ref{n-sol}), the number of players remaining in the
tournament, $M$, and the initial number of players $N$, are related by
$t\sim N/M$. Using (\ref{xt-scaling}), when $M$ players remain, the
typical rank is $x\sim (N/M)^{-(1-2q)}$. Substituting $M=1$, we find
that the typical rank of the winner decays algebraically with the
total number of players, as in (\ref{xn-scaling}), with the exponent
\begin{equation}
\label{beta-seq}
\beta_{\rm seq}=1-2q.
\end{equation}
Therefore, the smaller the tournament or the higher the upset
probability the weaker the winner, on average.  We note that due to
strong fluctuations, the master equation (\ref{f-eq}) is not
applicable when the number of players is of order one, and
consequently, our theoretical framework can not be used to obtain the
distribution of the tournament winner.

\noindent{\bf General Initial Distributions.} Our findings in the
case of uniform distributions suggest that the behavior of the
initial distribution in the $x\to 0$ limit governs the long time
asymptotics. Let us consider rank distributions with a power-law
behavior near the origin,
\begin{equation}
F_0(x)\simeq C\,x^{\mu+1},
\end{equation}
as $x\to 0$ with $\mu>-1$ so that the distribution is normalized.
The rank density then scales as \hbox{$f_0(x)\simeq
C(\mu+1)x^{\mu}$}. Since the rank $x$ decays with time, the term
$(1-F_0)(1+t)^{2q}$ in the denominator of (\ref{F-sol}) can be
replaced by $t^{2q}$ and similarly, the term $F_0(x)(1+t)$ can be
replaced by $Cx^{\mu+1}t$. Therefore, the scaling form
(\ref{Fxt-scaling}) becomes \hbox{$F(x,t)\to
t^{-1}\Phi\left(x\,t^{\frac{1-2q}{\mu+1}}\right)$}, with the scaling
function \hbox{$\Phi(z)=Cz^{\mu+1}/(1+Cz^{\mu+1})$}.  Thus the
typical player rank decays with time according to $x\sim
t^{-\frac{1-2q}{\mu+1}}$.  Similarly, the rank of the winner decays
with the number of players as in (\ref{xn-scaling}) with $\beta_{\rm
seq}=\frac{1-2q}{\mu+1}$.

Like the cumulative distribution, the density of players with given
rank also becomes self-similar asymptotically, \hbox{$f(x,t)\to
t^{\beta-1}\phi\big(x\,t^{\beta}\big)$} with
$\beta=\frac{1-2q}{\mu+1}$ and $\phi(z)=\Phi'(z)$. As noted earlier,
the shape of the distribution is preserved: $f(z)\sim z^{\mu}$ as
$z\to 0$. The large argument behavior is
\begin{equation}
\label{phi}
\phi(z)\sim z^{-\mu-2},
\end{equation}
as $z\to \infty$. The algebraic decay shows that the likelihood of
finding weak players in the tournament is appreciable. Surprisingly,
when initially most players are strong they can eliminate each
other, leading to an appreciable probability for weak players to
survive.

The scaling behavior (\ref{xn-scaling}) refers to the typical rank
of the winner. The algebraic tail (\ref{phi}) suggests that the
average rank may  scale differently than the typical rank. For
example, for compact uniform distributions ($\mu=0$), the average is
characterized by a logarithmic correction as in (\ref{xav}),
$\langle x_*\rangle \sim N^{-(1-2q)}\ln N$.

\noindent{\bf Parallel Dynamics.} Thus far, we addressed sequential
games with a single team eliminated at a time. However, actual
sports tournaments typically proceed via rounds of parallel play
with half of the teams eliminated in each round. We thus consider
such round-play tournaments with \hbox{$N=2^k$} players. Let
$g_N(x)$ be the normalized distribution of the rank of the winner
with $\int dx\, g_N(x)=1$ and let $G_N(x)=\int_0^x dy\, g_N(y)$ be
the corresponding cumulative distribution.

Consider first a tournament with $N=2$ players. Similar to
Eq.~(\ref{f-eq}), the rank distribution of the winner is
\begin{equation}
\label{f2-eq}
g_2(x)=2pg_1(x)[1-G_1(x)]+2qg_1(x)G_1(x).
\end{equation}
Integrating this equation, we arrive at an explicit expression for the
distribution of the rank of the winner
\hbox{$G_2(x)=2pG_1(x)+(1-2p)[G_1(x)]^2$}. Clearly, this nonlinear
recursion relation applies to every round of the tournament and
therefore,
\begin{equation}
\label{F-rec}
G_{2N}(x)=2pG_N(x)+(1-2p)[G_N(x)]^2.
\end{equation}
Iterating this equation starting with $G_1(x)$, we obtain explicit
expressions for the distribution of the winner for $N=2,4,8,\ldots$
Explicit expressions can be obtained for the extreme cases of
deterministic competitions ($q=0$) where $1-G_N(x)=[1-G_1(x)]^N$ and
random competitions ($q=1/2$) where $G_N(x)=G_1(x)$.

Let us restrict our attention to uniform initial distributions,
$G_1(x)=x$ for $x\leq 1$. For small-$x$, we may neglect the
nonlinear term in (\ref{F-rec}) and then, $G_2(x)\simeq (2p)\,x$,
$G_4(x)\simeq (2p)^2\,x$, and in general
\begin{equation}
\label{F2k-sol}
G_{2^k}(x)\simeq (2p)^k\,x.
\end{equation}
To obtain the asymptotic behavior, we substituting \hbox{$k=\frac{\ln
N}{\ln 2}$} into (\ref{F2k-sol}) and then $G_N(x)\simeq N^\beta\,x$
with \hbox{$\beta=1+\frac{\ln p}{\ln 2}$}. Therefore, the cumulative
distribution of the rank of the winner follows the scaling form
\begin{equation}
\label{FN-scaling}
G_N(x)\to\Psi\left(x\,N^\beta\right)
\end{equation}
when $N\to \infty$.  The scaling function is linear,
\hbox{$\Psi(z)\simeq z$}, in the limit $z\to 0$, reflecting
that the extremal statistics are invariant under the competition
dynamics.

\begin{figure}[t]
\includegraphics*[width=0.425\textwidth]{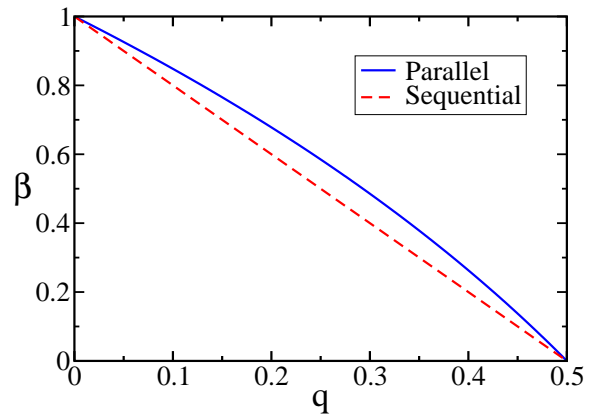}
\caption{The decay exponent $\beta$ versus the upset probability $q$.
  Shown are the values for the sequential case (\ref{beta-seq}) and
  the parallel case (\ref{beta-par}).}
\label{beta-fig}
\end{figure}

The scaling form (\ref{FN-scaling}) shows that the rank of the
tournament winner decays algebraically with the tournament size as in
(\ref{xn-scaling}). Surprisingly, the decay exponent
\begin{equation}
\label{beta-par}
\beta_{\rm par}=1+\frac{\ln (1-q)}{\ln 2}
\end{equation}
for parallel dynamics, differs from the decay exponent
(\ref{beta-seq}) for sequential dynamics.  The two exponents
coincide in the extreme cases, $\beta(0)=1$ and $\beta(1/2)=0$. The
inequality $\beta_{\rm par}\geq \beta_{\rm seq}$ (figure
\ref{beta-fig}) shows that parallel play benefits the strong
players. Indeed, in sequential play weak players may survive by
being idle. The source of this discrepancy is fluctuations in the
number of games. In sequential dynamics, the number of games played
by each player is variable while in parallel dynamics the number of
games is fixed.

Substituting the scaling form (\ref{FN-scaling}) into the recursion
(\ref{F-rec}), the scaling function obeys the nonlinear-nonlocal
equation
\begin{equation}
\label{Psi-eq}
\Psi(2pz)=2p\,\Psi(z)+(1-2p)\Psi^2(z).
\end{equation}
The boundary condition are $\Psi(0)=0$ and $\Psi(\infty)=1$.  An
exact solution is feasible only when there are no upsets:
$\Psi(z)=1-e^{-z}$ for $q=0$. Otherwise, we perform an asymptotic
analysis. As shown above, the small-$z$ behavior is generic,
$\Psi(z)\simeq z$. At large arguments, we write $U(z)=1-\Psi(z)$ and
since $U\ll 1$, we can neglect the nonlinear terms and then
$U(2pz)=2qU(z)$. This implies the algebraic decay $U(z)\sim z^{(\ln
2q)/(\ln 2p)}$. As a result, the likelihood of finding weak winners,
$g_N(x)\to N^\beta\psi\left(xN^\beta\right)$ with
$\psi(z)=\Psi'(z)$, decays algebraically
\begin{equation}
\label{psi-tail} \psi(z)\sim z^{\frac{\ln 2q}{\ln 2p}-1}
\end{equation}
as $z\to\infty$. This algebraic behavior is very different from the
exponential decay $\psi(z)=e^{-z}$ for deterministic games.  In
contrast to sequential play, the exponent depends on the upset
probability.  This large likelihood of finding weak winners reflects
that the number of games played by the tournament winner scales
logarithmically with the number of teams. For example, as $N=2^k$,
the likelihood that the weakest player wins, $q^k=N^{\ln q/\ln 2}$,
is appreciable as it decays only algebraically with $N$.

\begin{figure}[t]
\includegraphics*[width=0.425\textwidth]{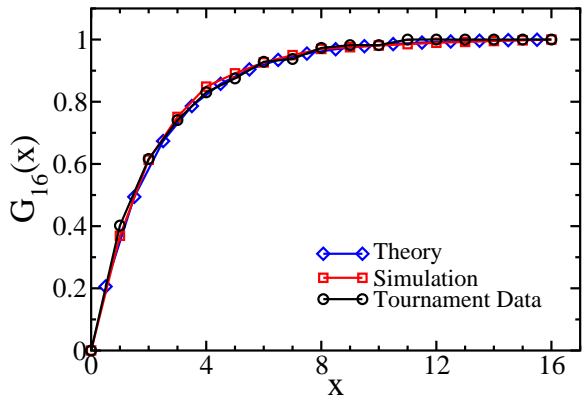}
\caption{The cumulative distribution of the rank of the group winner
$G_{16}(x)$. The empirical distribution for college basketball
(circle) is compared with Monte Carlo simulations (squares), and the
parallel dynamics theory (diamonds).} \label{dist-fig}
\end{figure}

\noindent{\bf Empirical Study.} To test our theoretical approach, we
studied the US men's NCAA college basketball national championship
where 64 teams are divided into 4 groups of 16, with teams in each
group ranked 1 (best) to 16 (worst). The winner of each group
advances to the ``final four''. As in the parallel dynamics, half of
the teams are eliminated in each round. The schedule, however, is
not random: the games are arranged so that if there are no upsets
the bottom half is eliminated in each round. We analyzed the results
of all 1680 games since this format was established (1979-2006)
\cite{data}. We calculated the cumulative rank distribution of the
team advancing to the final four, $G_{16}(x)$, with
$x=1,2,\ldots,16$ (figure \ref{dist-fig}). Additionally, we measured
the upset frequency $q=0.275$ by counting the number of games won by
the underdog \cite{bvr1}.

To compare with the theoretical model, we simulated the NCAA
tournament schedule in which the lower-ranked team wins with upset
probability $q$. The parameter $q$ was treated as a tunable variable,
and we present results for the value that best matched the empirical
data. The simulation results produce a rank distribution that agrees
well with the empirical findings (figure \ref{dist-fig}). The fitted
upset probability $q=0.22$ is close to the observed frequency.
Alternatively, we modeled the data by iterating (\ref{F-rec}) starting
with the uniform distribution $G_1(x)=x/16$ using a fitted upset
probability of $q=0.175$ (the theory assumes a random schedule and an
approximate uniform distribution). We thus found that the competition
model has predictive power that quantitatively captures empirical rank
distributions, and enables estimates of upset frequencies from
observed rank distributions.

In summary, we studied dynamics of single-elimination tournaments,
in which there is a finite probability for a lower-ranked player to
upset a higher-ranked player.  We obtained an exact solution for the
distribution of player ranks for arbitrary initial conditions.
Generally, the likelihood of upset winners is relatively large since
the tail of the distribution function decays algebraically with
rank. The characteristic rank of the winning player decays
algebraically with the number of players and the larger the upset
probability, the slower this decay (small tournaments are more
likely to produce a surprise winner). Different decay exponents are
found for sequential and parallel play with the latter generally
larger (weak players fare better by avoiding competition). We
demonstrated the utility of this model using college basketball
results.

Extreme properties of the initial distribution fully governs the
asymptotic behavior.  In the long time limit, the player
distribution becomes self-similar.  Both the form of the scaling
distribution and the time dependence of the characteristic rank
depend only on the small-$x$ behavior of the initial distribution. A
similar phenomenology where extremal statistics governs long-time
asymptotics was found in studies of clustering in traffic flows
\cite{bkr} and species abundance in biological evolution
\cite{jk,smd}.

\noindent{\bf Acknowledgments.} We thank the Isaac Newton Institute
for Mathematical Sciences (Cambridge, England) and the Asia Pacific
center for Theoretical Physics (Pohang, South Korea) for their
hospitality. We acknowledge financial support from DOE grant
DE-AC52-06NA25396 and NSF grant DMR0535503.

\end{document}